\documentclass
[superscriptaddress,secnumarabic,amssymb,amsmath,nobibnotes,aps,prd,showkeys,showpacs,nofootinbib,onecolumn]{revtex4}%
\usepackage{graphicx}
\usepackage{epsf}
\usepackage{bm}
\usepackage{amsmath}
\usepackage{amsfonts}
\usepackage{amssymb}%
\providecommand{\U}[1]{\protect\rule{.1in}{.1in}}

\newcommand{\be}{\begin{equation}}
\newcommand{\ee}{\end{equation}}

\newcommand{\mincir}{\raise
-3.truept\hbox{\rlap{\hbox{$\sim$}}\raise4.truept\hbox{$<$}\ }}
\newcommand{\magcir}{\raise
-3.truept\hbox{\rlap{\hbox{$\sim$}}\raise4.truept\hbox{$>$}\ }}

\begin{document}
\title{Dynamical analysis and cosmological viability of varying $G$ and $\Lambda$ cosmology}
\author{Andronikos Paliathanasis}
\email{anpaliat@phys.uoa.gr}
\affiliation{Instituto de Ciencias F\'{\i}sicas y Matem\'{a}ticas, Universidad Austral de
Chile, Valdivia, Chile}
\affiliation{Department of Mathematics and Natural Sciences, Core Curriculum Program,
Prince Mohammad Bin Fahd University, Al Khobar 31952, Kingdom of Saudi Arabia}
\affiliation{Institute of Systems Science, Durban University of Technology, PO Box 1334,
Durban 4000, Republic of South Africa}

\begin{abstract}
The cosmological viability of varying $G\left(  t\right)  $ and $\Lambda
\left(  t\right)  $ cosmology is discussed by determining the cosmological
eras provided by the theory. Such a study is performed with the determination
of the critical points while stability analysis is performed. The application
of Renormalization group in the ADM formalism of General Relativity provides a
modified second-order theory of gravity where varying $G\left(  t\right)  $
plays the role of a minimally coupled field, different from that of
Scalar-tensor theories, while $\Lambda\left(  t\right)  =\Lambda\left(
G\left(  t\right)  \right)  $ is a potential term. We find that the theory
provides two de Sitter phases and a tracking solution. In the presence of
matter source, two new critical points are introduced, where the matter source
contributes to the universe. One of those points describes the $\Lambda$CDM
cosmology and in order for the solution at the point to be cosmologically
viable, it has to be unstable. Moreover, the second point, where matter
exists, describes a universe where the dark energy parameter for the equation
of state has a different value from that of the cosmological constant.

\end{abstract}
\maketitle
\date{\today}

\section{Introduction}

The detailed analysis of the cosmological data over the last years supports
the assumptions that the universe is spatially flat, it has been through an
inflation phase in the past prior to the radiation dominated era, and that,
currently. the universe is in a second acceleration epoch
\cite{dataacc2,data1,data3,data4,planck2015}. The acceleration phase of the
universe has been attributed to a matter source in the gravitational field
equations which has an equation of state parameter with a negative value. The
nature of this exotic matter source has led to the dark energy problem.

In the literature one can find various proposals/models to solve the dark
energy problem. These proposals can be categorized in two different families;
more specifically, in these where in the context of Einstein's General
Relativity, an energy momentum tensor is introduced to explain the
acceleration phases \cite{ar1,ar2,ar3,ar4,ar5,ar12,ar6,ar7,ar8,ar9,ar11}, and
these in which the Einstein-Hilbert action is modified, leading to the
so-called modified/alternative theories of gravity, such that the origin of
the acceleration to correspond to the gravitational theory, for instance, see
\cite{fr0,fr1,fr2,fr3,fr4,fr5,fr6,fr10,fr11,fr12,fr13,fr14,fr15} and
references therein.

A common feature for some of the modified theories of gravity is that Newton's
constant $G,$ is varying; and it is a varying parameter. For instance, in
Brans-Dicke theory and in $f\left(  R\right)  $-gravity someone can define the
effective parameters $G_{eff}=G\phi^{-1}$ and $G_{eff}=Gf^{\prime}\left(
R\right)  $ respectively~\cite{fr0,fr1,jdb}. Dealing with fundamental
constants in physics as parameters is the main concept of the renormalization
group \cite{ref0,ref0a,ref0b}. Reuter and Weyer in \cite{ref1} inspired by the
property that Brans-Dicke action modify $G$, reconstructed the Brans-Dicke
action with the use of the renormalization group in General Relativity, by
assuming that $G$ and $\Lambda$ (the cosmological constant) are varying.
Various alternative gravitational theories
\cite{aref0,aref1,aref1a,aref2,aref3,aref4,aref5,aref6a,aref6b,aref6c,aref6d,aref6e,aref6f,aref01,aref02,areff}
have been modified by the renormalization group in cosmological systems as
also in strong gravitational systems
\cite{aref6,aref7,aref8,aref9,aref10,aref11}.

A study which provides important analytical information about the existence of
cosmological epochs (such as matter dominated era, acceleration phase and
others) and the stability of those epochs is the analysis of critical points
of the gravitational filed equations \cite{bia1,bia2}. In the dark energy
models, the analysis of the critical point provides results for the evolution
of the universe \cite{copeland} and the viability of each model being studied
\cite{dn4}. For some extended applications of the critical point analysis in
modified theories of gravity, we refer the reader to
\cite{dn2,dn3,dn6,dn7,dn8,dn12,dn12a,dn12b} and references therein.

We are interested in the dynamical analysis of the gravitational field
equations which follows from the renormalization group in the ADM Lagrangian
of General Relativity as described in \cite{alfio1}. Specifically, in
\cite{alfio1} the authors assumed that $G$ and $\Lambda$~are varying
parameters such that new degrees of freedom are introduced. The theory,
remains of second-order and the variable $G$ can be seen as a scalar field
coupled to gravity, but different from that of Brans-Dicke or from the
scalar-tensor theory. The reason for the latter lies in the starting point for
the application of the renormalization group. This is the ADM Lagrangian and
not the Einstein-Hilbert action as in \cite{ref1}. Some\ exact solutions for
that specific modified gravitational theory can be found in
\cite{alfio2,alfio3}. Cosmological constraints and comparison with the
$\Lambda$CDM model are given in \cite{ester} where it was found that for this
specific variable $G,~\Lambda$ cosmology is compatible with some of the
observational data and can explain the late acceleration phase of the universe.

More specifically, in this work, we study the existence of critical points in
varying $G,~\Lambda$cosmology \cite{alfio1} in order to explore the possible
cosmological eras provided by the theory. We define new dimensionless
variables and in terms of the $H-$normalization \cite{copeland} we study the
critical points of the cosmological model. Because the resulting field
equations of \cite{alfio1} have similarities with Scalar-tensor theories, our
analysis can be compared with the analysis performed for the Brans-Dicke
theory in \cite{dn8}. However, as we shall see, there are essential
differences with the Scalar-tensor theories. The plan of the paper follows.

In Section \ref{sec2} we present the model of our consideration which belongs
to the family of varying $G$ and $\Lambda$ cosmology. Section \ref{sec3}
includes the main material of our analysis where the analysis of the critical
points for dimensionless variables and in the $H$-normalization is discussed.
Our discussion of the results is given in Section \ref{con}, where we also
draw our conclusions.

\section{Field equations in varying $G$ and $\Lambda$ cosmology}

\label{sec2}

In the ADM formalism of General Relativity, Bonanno et al. \cite{alfio1} after
the application of the renormalization group, proposed the following
modification for the ADM\ Lagrangian of General Relativity,%
\begin{align}
S  &  =S_{m}+\frac{1}{16\pi}\int\frac{N\sqrt{h}}{G}\left(  K_{ij}K^{ij}%
-K^{2}+R^{\ast}-2\Lambda\left(  G\right)  \right)  d^{3}x+\nonumber\\
&  +\frac{\mu}{16\pi}\int\frac{N\sqrt{h}}{G}\left(  N^{-2}\left(
G_{,0}\right)  ^{2}-2\frac{N^{i}}{N^{2}}G_{,0}G_{,i}-\left(  h^{ij}%
-\frac{N^{i}N^{j}}{N^{2}}\right)  G_{,i}G_{,j}\right)  d^{3}x, \label{gl1}%
\end{align}
where $S_{m}$ describes the Action Integral of the matter source, and
$G,~\Lambda\left(  G\right)  $ are varying.

Furthermore, the line element of the background metric in the ADM formalism is
expressed as \cite{adm1}%
\begin{equation}
ds^{2}=-\left(  N^{2}-N_{i}N^{i}\right)  dt^{2}+2N_{i}dtdx+h_{ij}dx^{i}%
dx^{j}~,~i,j=1,2,3, \label{gl2}%
\end{equation}
in which $N$ denotes the lapse function, $N^{i}$ are the components of the
shift vector, $h_{ij}$ is the metric tensor three-dimensional surface
\cite{adm1,adm2a}. $K_{ij}$ denotes the extrinsic curvature and $R^{\ast}$ the
curvature of the three-dimensional surface with metric tensor $h_{ij}$.

In the special consideration of a spatially flat isotropic and homogeneous
universe, line element (\ref{gl2}) is that of the
Friedmann-Lema\^{\i}tre-Robertson-Walker (FLRW) geometry, that is,
\begin{equation}
ds^{2}=-N^{2}dt^{2}+a^{2}\left(  t\right)  \left(  dx^{2}+dy^{2}%
+dz^{2}\right)  . \label{gl3}%
\end{equation}

Therefore, the Action Integral (\ref{gl1}) is simplified and the following
point-like Lagrangian can be extracted\cite{alfio1}%
\begin{equation}
\mathcal{L}\left(  N,a,\dot{a},G,\dot{G}\right)  =\frac{1}{N}\left(  -\frac
{3}{G}a\dot{a}^{2}+\frac{\mu}{2G}a^{3}\left(  \frac{\dot{G}}{G}\right)
^{2}\right)  -Na^{3}V\left(  G\right)  +Na^{3}\rho_{m}, \label{gl4}%
\end{equation}
where $\Lambda\left(  G\right)  =GV\left(  G\right)  $ and $\rho_{m}$ presents
the contribution of the matter source.

For the matter source, we assume that it describes a dust fluid which
attributes the dark matter source of the universe and it is minimally coupled
to gravity, that is $p_{m}=0$ and $\rho_{m}=8\pi\rho_{m0}a^{-3}$. At this
point, it is important to mention that we have assumed the comoving observer
$u^{\mu}=\frac{1}{N}\delta_{0}^{a}$, such that $u^{\mu}u_{\mu}=-1$.

Lagrangian (\ref{gl4}) describes a second-order theory with degrees of freedom
$\left\{  N,a,G\right\}  $. Specifically, the variation with respect to the
lapse function provides the constraint equation, while two second-order
equations follow from the variation with respect to the rest parameters
$a\left(  t\right)  $ and $G\left(  t\right)  $. Parameter $\mu$ denotes the
interaction; its value is unknown and it is a dimensionless parameter
\cite{alfio1}. It is analogue to the Brans-Dicke parameter. Furthermore, it is
important parameter $\mu$ to be nonzero in order the field equations to admit
nontrivial solutions \cite{alfio1}.

Variation with respect to the dependent variables $\left\{  N,a,G\right\}  $
in Lagrangian (\ref{gl4}) derives the modified gravitational field equations
\cite{alfio1,alfio2,alfio3}
\begin{equation}
-\frac{3}{G}a\dot{a}^{2}+\frac{\mu}{2G}a^{3}\left(  \frac{\dot{G}}{G}\right)
^{2}+a^{3}V\left(  G\right)  =-8\pi\rho_{m0}, \label{gl5}%
\end{equation}%
\begin{equation}
\ddot{a}+\frac{1}{2a}\dot{a}^{2}-\dot{a}\frac{\dot{G}}{G}+\frac{\mu}{4}%
a\frac{\dot{G}^{2}}{G^{2}}-\frac{1}{2}GaV=0, \label{gl6}%
\end{equation}%
\begin{equation}
\ddot{G}-\frac{2}{\mu}G\left(  \frac{\dot{a}}{a}\right)  ^{2}+3\frac{\dot{a}%
}{a}\dot{G}-\frac{3}{2}\frac{\dot{G}^{2}}{G}+\mu G^{3}V_{,G}=0, \label{gl7}%
\end{equation}
where without loss of generality we have set the lapse function to be
constant, i.e. $N\left(  t\right)  =1$.

Equations (\ref{gl5}), (\ref{gl6}) are the modified Friedmann's equations,
while equation (\ref{gl7}) is the corresponding \textquotedblleft
Klein-Gordon\textquotedblright\ equation for the \textquotedblleft
field\textquotedblright\ $G\left(  t\right)  $. It is important to mention
that $G\left(  t\right)  $ does not belong to the family of scalar-tensor
theories \cite{farbook}.

An equivalent way to write the field equations is with the use of the Hubble
function $H=\frac{\dot{a}}{a}$, that is,%
\begin{equation}
3H^{2}=G_{eff}\left(  \rho_{m}+\rho_{G}\right)  ,\label{gl8}%
\end{equation}%
\begin{equation}
2\dot{H}+3H^{2}=-G_{eff}p_{G},\label{gl9}%
\end{equation}
and%
\begin{equation}
\frac{\ddot{G}}{G}-\frac{2}{\mu}H^{2}+3\left(  \frac{\dot{G}}{G}\right)
H-\frac{3}{2}\left(  \frac{\dot{G}}{G}\right)  ^{2}+\mu G^{2}V_{,G}%
=0,\label{gl.10}%
\end{equation}
from where $\rho_{G}\left(  t\right)  ,~p_{G}\left(  t\right)  $ denotes the
energy density and the pressure component related to the field $G\left(
t\right)  $, as follows%
\begin{equation}
\rho_{G}=\frac{1}{2}\mu\left(  \frac{\dot{G}}{G}\right)  ^{2}+GV\left(
G\right)  ,\label{gl10}%
\end{equation}%
\begin{equation}
p_{\phi}=-2H\frac{\dot{G}}{G^{2}}+\frac{\mu}{2}\frac{\dot{G}^{2}}{G^{3}%
}-V\left(  G\right)  ,\label{gl11}%
\end{equation}
while $G_{eff}=8\pi G\left(  t\right)  $ is the effective time-varying
Newton's \textquotedblleft constant\textquotedblright. Furthermore, from
(\ref{gl5}) and (\ref{gl6}) we observe that the Einstein field equations are
recovered provided that $G=G_{0}$, i.e. a constant, and $\Lambda\left(
G_{0}\right)  $ satisfies the equation%
\begin{equation}
G_{0}V_{,G}\left(  G_{0}\right)  -V\left(  G_{0}\right)  =0.
\end{equation}
The latter equation is always true for $V\left(  G\right)  =V_{0}G$, that is,
$\Lambda\left(  G\right)  \simeq G^{2}$. 

In the following section the detailed analysis for the critical points of the
field equations (\ref{gl5})-(\ref{gl7}) is presented

\section{Dynamical analysis}

\label{sec3}

In this Section, we study the existence and the stability of critical/fixed
points for the gravitational field equations. In order to perform such an
analysis we define dimensionless variables in the $H-$normalization, see
\cite{copeland}. The novelty of these coordinates is that any critical point
corresponds to a power-law scale factor, i.e. $a\left(  t\right)  =a_{0}t^{p}$
or to a de Sitter universe with exponential scale factor, i.e. $a\left(
t\right)  =a_{0}e^{H_{0}t}$.

\subsection{Dimensionless variables}

We continue by defining the new dimensionless variables in the $H-$%
normalization approach,
\begin{equation}
x=\frac{\mu}{6H^{2}}\left(  \frac{\dot{G}}{G}\right)  ^{2}~~,~~y=\frac
{GV\left(  G\right)  }{3H^{2}}~~,~~\Omega_{m}=G\frac{\rho_{m}}{3H^{2}},
\label{gl12}%
\end{equation}
while the constraint equation (\ref{gl5}) takes the algebraic form%
\begin{equation}
1-x-y=\Omega_{m}, \label{gl13}%
\end{equation}
from where it follows that since $\Omega_{m}\in\left[  0,1\right]  $, then
$0\leq1-x-y\leq1$. Parameters are not necessarily positive. The sign of $x$
depends on the interaction parameter $\mu$, while the sign of variable $y$
depends on the sign of the varying $\Lambda\left(  G\left(  t\right)  \right)
$. \ Moreover, the energy density of the field $G$ is defined as $\Omega
_{G}=x+y$.

Consider now the new independent parameter $\tau=\ln a$; then second-order
differential equations (\ref{gl6}) and (\ref{gl7}) can be written as the
first-order ordinary differential equations%
\begin{equation}
\frac{dx}{d\tau}=-3x\left(  1-x+y\right)  -\sqrt{\frac{6x}{\mu}}\left(
1-x-\lambda y\right)  , \label{gl14}%
\end{equation}%
\begin{equation}
\frac{dy}{d\tau}=3y\left(  1+x-y\right)  +\sqrt{\frac{6x}{\mu}}y\left(
1-\lambda\right)  , \label{gl15}%
\end{equation}%
\begin{equation}
\frac{d\lambda}{d\tau}=-\sqrt{\frac{6x}{\mu}}\left(  1-\lambda\Gamma\left(
\lambda\right)  \right)  , \label{gl16}%
\end{equation}
in which the new parameter $\lambda$ and function $\Gamma\left(
\lambda\right)  $ are defined as
\begin{equation}
\lambda=G\left(  \ln V\right)  _{,G}~~\text{and~~}\Gamma\left(  \lambda
\right)  =\frac{V_{,GG}V}{\left(  V_{,G}\right)  ^{2}}-1. \label{gl17}%
\end{equation}

As far as the equation of state parameter for the dark energy fluid term is
concerned, from the definition of (\ref{gl10}) and (\ref{gl11}) with the use
of the variables (\ref{gl12}) we calculate%
\begin{equation}
w_{G}\left(  x,y\right)  =-\left(  \frac{x-y}{x+y}\right)  -2\sqrt{\frac
{2}{3\mu}}\frac{\sqrt{x}}{x+y}. \label{gl18}%
\end{equation}

The deceleration parameter,~$q=-1-\frac{\dot{H}}{H^{2}}~$is expressed as%
\begin{equation}
q\left(  x,y\right)  =\frac{1}{2}\left(  1+3\left(  x-y\right)  -2\sqrt
{\frac{6x}{\mu}}\right)  , \label{gl18a}%
\end{equation}
and the equation of state parameter for the total fluid is derived to be%
\begin{equation}
w_{tot}\left(  x,y\right)  =\left(  x-y\right)  -2\sqrt{\frac{2x}{3\mu}}.
\label{gl.18b}%
\end{equation}

The dynamical system (\ref{gl14})-(\ref{gl16}) in general has dimension three.
However, the dimension of the system is reduced by one in the vacuum, with the
use of the algebraic equation (\ref{gl13}). Another possible case where the
dimension is reduced is when $\lambda$ is an identical constant, that is
$\lambda=\lambda_{0}$, which corresponds to the power-law potential $V\left(
G\right)  =V_{0}G^{\lambda_{0}},$ that is, $\Lambda\left(  G\right)
=V_{0}G^{\lambda_{0}+1}$, with $\lambda_{0}\neq0.$

\subsection{Critical points in the vacuum}

Consider the vacuum scenario, $\Omega_{m}=0$, where from the constraint
equation (\ref{gl13}) it follows $y=1-x$. Therefore, the reducing dynamical
system is%
\begin{equation}
\frac{dx}{d\tau}=-6\left(  1-x\right)  x-\sqrt{\frac{6x}{\mu}}\left(
1-x\right)  \left(  1-\lambda_{0}\right)  , \label{gl19}%
\end{equation}%
\begin{equation}
\frac{d\lambda}{d\tau}=-\sqrt{\frac{6x}{\mu}}\left(  1-\lambda\Gamma\left(
\lambda_{0}\right)  \right)  , \label{gl20}%
\end{equation}
while, as we have discussed before for a power-law potential in which
$\lambda=const.,$ the latter dynamical system reduced to the one-dimensional
system (\ref{gl19}).

We continue by assuming two special forms for the potential, (a) power-law
potential $V_{\left(  a\right)  }\left(  G\right)  =V_{0}G^{\lambda_{0}}$,
where $\lambda_{0}$ is a constant, and (b) exponential potential $V_{\left(
b\right)  }\left(  G\right)  =V_{0}\exp\left(  \lambda_{0}G\right)  $, such
that $\Gamma\left(  \lambda\right)  =\left(  \lambda_{0}\right)  ^{-1}$ is a
constant parameter. The critical points of these two potentials are the only
physically different possible points. It is possible for another potential the
dynamical system (\ref{gl19}), (\ref{gl20}) to admit more critical points from
the potentials $V_{\left(  a\right)  },~V_{\left(  b\right)  }$; however, the
physical properties will be on that of the points of potentials $V_{\left(
a\right)  }~$and$~V_{\left(  b\right)  }\,.$

\subsubsection{Power-law potential}

Consider the power-law potential, $V_{\left(  a\right)  }\left(  G\right)
=V_{0}G^{\lambda_{0}}$, then the equilibrium points of equation (\ref{gl19})
are%
\[
P_{1}:x=0~,~P_{2}:~x=1\text{ and }P_{3}:x=\frac{\left(  1-\lambda_{0}\right)
^{2}}{6\mu}.
\]
where point $P_{3}$ depends on the value of the constant $\lambda_{0}.~$

Below we discuss the physical properties and the stability of each point.

\begin{figure}[ptb]
\includegraphics[height=6.5cm]{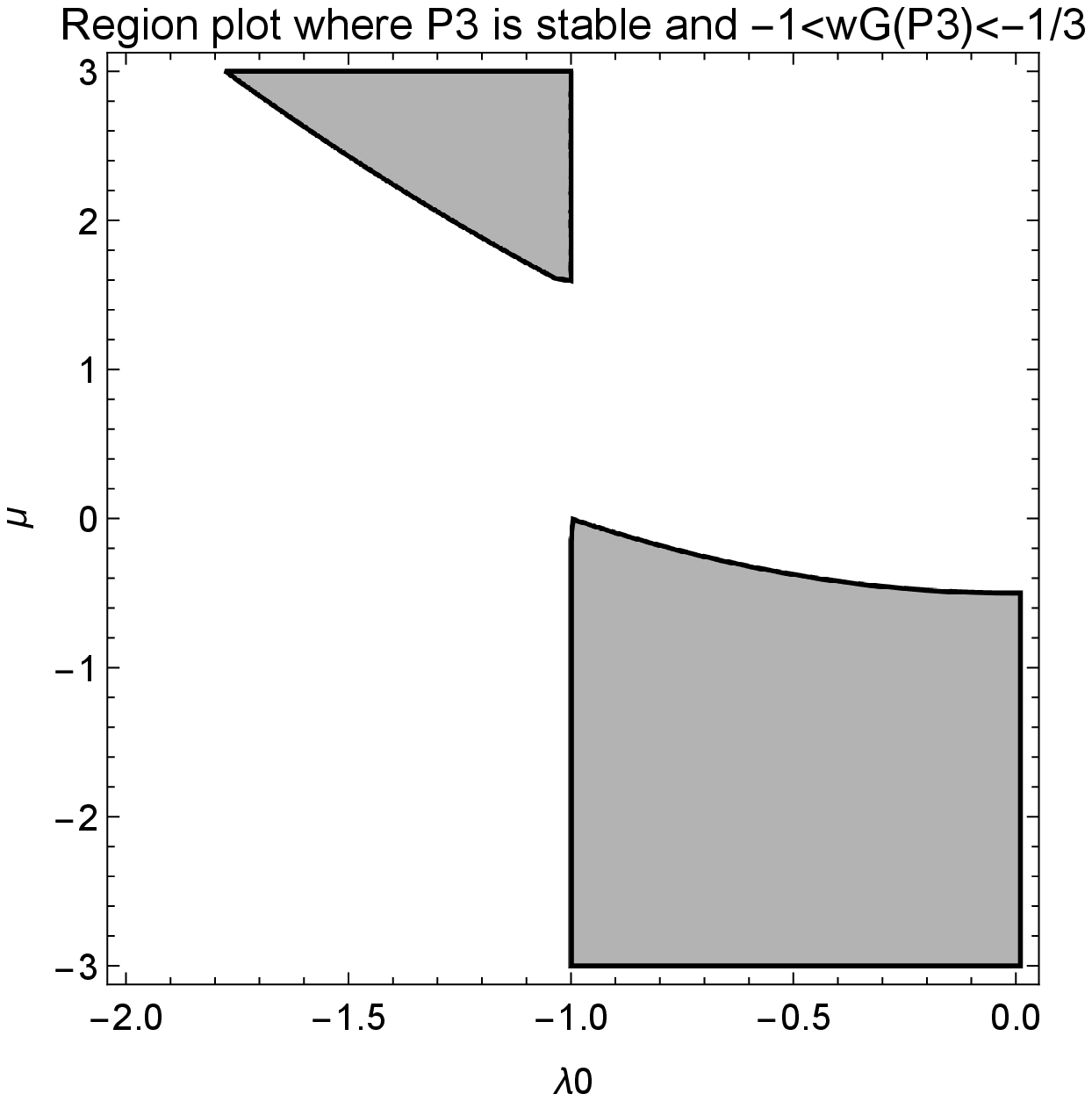}
\includegraphics[height=6.5cm]{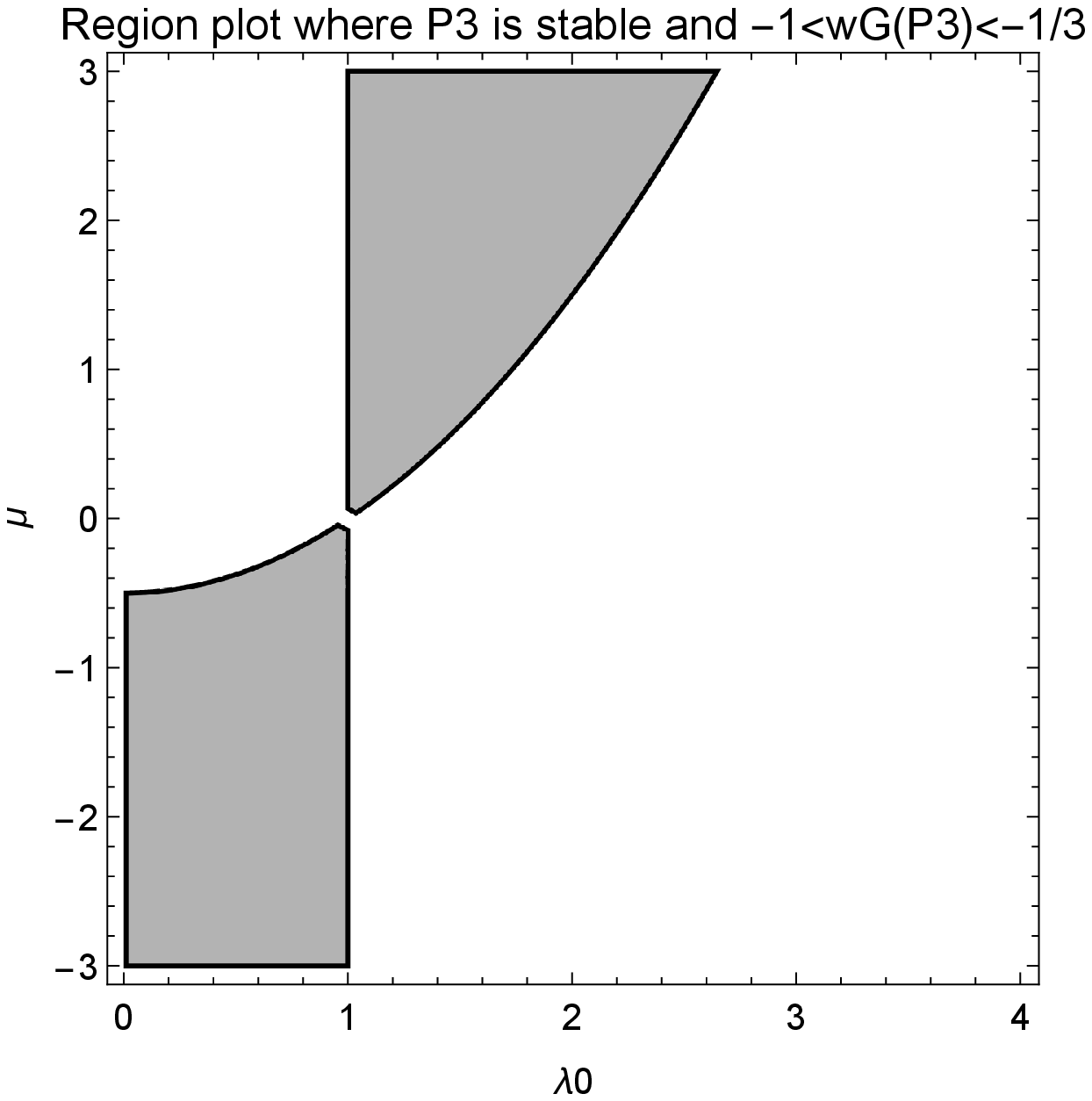}
\caption{Region plot in the space $\left\{  \mu,\lambda_{0}\right\}
~$for$~0<\left\vert \mu\right\vert <3\,\ and$ $-2<\lambda_{0}<0~$(Left
Fig.)$,~0<\lambda_{0}<4~$(Right Fig.), where point $P_{3}$ is stable and
describes an accelerated universe with$-1\leq w_{G}\left(  P_{3}\right)
<-\frac{1}{3}$.}%
\label{p3eos}%
\end{figure}

\begin{itemize}
\item Point $P_{1}$ corresponds to the epoch in which the potential $V\left(
G\right)  $ dominates the universe and the $\dot{G}=0$, that is, $V\left(
G\right)  $ is the cosmological constant. Hence, $w_{tot}\left(  P_{1}\right)
=w_{G}\left(  P_{1}\right)  =-1$, and describes $P_{1}$ is a de Sitter point,
which can describe the past inflationary epoch when $P_{1}$ is unstable; or it
can be a future attractor in the evolution of the universe when~$P_{1}$ is a
stable point. The stability of the point depends on the value of the power
$\lambda_{0}$. In particular for values of $\mu\,,~$in which $\mu x$ is
positive close to the limit $x\rightarrow0,$ point $P_{1}$ is unstable for
$\lambda_{0}>1$, while for $\lambda_{0}\leq1$ the eigenvalue has a negative
limit and the point is stable. On the other hand when $\mu$ is negative close
to the limit $x\rightarrow0$; for instance $\ $for $\mu<0$ and $x\rightarrow
0^{+}$~point $P_{1}~$describes a stable spiral.

\item Point $P_{2}$ corresponds to the epoch in which the kinetic term
dominates the universe and $\rho_{G}\left(  P_{2}\right)  =\frac{1}{2}%
\mu\left(  \frac{\dot{G}}{G}\right)  ^{2}$. The equation of state parameter is
calculated to be $w_{tot}\left(  P_{2}\right)  =w_{G}\left(  P_{2}\right)
=1-2\sqrt{\frac{2}{3\mu}}$, which is real for positive values of the parameter
$\mu.$ The point describes an accelerated universe,~i.e. $w_{G}\left(
P_{2}\right)  <-\frac{1}{3}$, for values of $\mu$ in the range $0<\mu<\frac
{3}{2}$, while $w_{G}\left(  P_{2}\right)  \geq-1$ for $\frac{2}{3}\leq
\mu<\frac{3}{2}$. Hence, the scale factor at the point $P_{2}$ is exponential
for $\mu=\frac{2}{3}$, and power-law for other values of $\mu.$ For $\mu
=\frac{8}{3}$ and $\mu=6\,\ $point $P_{2}$ corresponds to eras where the field
$G$ behaves like dust or radiation fluids respectively. It is important to
mention that there is not any finite value of $\mu$ such that the geometric
matter source, $\rho_{G}$, has the equation of state parameter of the stiff
fluid. Finally, point $P_{2}$ is stable for all the values of $\mu$ which are
defined where $\lambda_{0}<1+\sqrt{6\mu}$.

\item Point $P_{3}$ exists when $\lambda\neq1$, from where we calculate that
$w_{tot}\left(  P_{3}\right)  =w_{G}\left(  P_{3}\right)  =-1-\frac{2\left(
1-\lambda_{0}\right)  }{3\mu}+\frac{\left(  1-\lambda_{0}\right)  ^{2}}{3\mu}%
$. \ The stability of the point depends on the parameters $\mu~$and $\lambda$;
specifically, the point is stable when $\mu<0$ or $\mu>\frac{\left(
1-\lambda_{0}\right)  ^{2}}{6}$. \ The point $P_{3}$ describes acceleration
for ranges of the free parameters $\lambda_{0},~\mu$ in which: (i) $\left\{
\lambda_{0}<1~,~\lambda_{0}>3\right\}  ,~\mu<0$ or $2\mu>\left(  2-\lambda
_{0}\right)  ^{2}-1;~$(ii) $1<\lambda_{0}\leq3$, $\mu>0~$or $2\mu>\left(
2-\lambda_{0}\right)  ^{2}-1.$ Finally, for $\lambda=3$, $P_{3}$ describes a
de Sitter universe. Thus, it is clear that except from the coordinates of
point $P_{3}$, the eigenvalue of the point depends on the constant
$\lambda_{0}$ which, in general, can take any value except for zero. In Fig.
\ref{p3eos} the surface where point $P_{3}$ is stable and describes an
accelerated universe such that $w_{G}\left(  P_{3}\right)  \in\lbrack-1,1/3)$
is plotted in the space of the parameters $\left\{  \mu,\lambda_{0}\right\}  $
for $-2<\lambda_{0}<0,~0<\lambda_{0}<4,$~and $0<\left\vert \mu\right\vert <3.$
\end{itemize}

\subsubsection{Exponential Potential}

In the case of the exponential potential $V_{\left(  b\right)  }\left(
G\right)  $, where $\Gamma\left(  \lambda\right)  $ is constant, i.e.
$\Gamma\left(  \lambda\right)  =\left(  \lambda_{0}\right)  ^{-1}$, the
stationary points $Q\left(  x,\lambda\right)  $ for the dynamical system
(\ref{gl19}) and (\ref{gl20}) are%
\[
Q_{1}=\left(  0,\lambda_{0}\right)  ~~,~~Q_{2}=\left(  1,\lambda_{0}\right)
~~,~~Q_{3}=\left(  \frac{\left(  1-\lambda_{0}\right)  ^{2}}{6\mu},\lambda
_{0}\right)  ~,
\]%
\[
Q_{4}=\left(  \frac{1}{6\mu},0\right)  ~\text{~and~~}Q_{5}=\left(  1,0\right)
.
\]

Points $Q_{1-3}$ are the points $P_{1-3}$ for the power-law potential in which
$\lambda=\lambda_{0}~,~\lambda_{0}\neq0,$ and have the same physical
properties. Points $Q_{4}$ and $Q_{5}$ are new points. The discussion on the
physical properties and the stability of the critical points follows.

\begin{itemize}
\item Point $Q_{1}$ actually describes invariant manifold of the dynamical
system rather than a stationary point in the space $\left\{  x,\lambda
\right\}  $. Any point on the line, $x=1$, has the same physical properties
with $P_{1}$, that is the universe is dominated by the potential $V\left(
G\right)  $, which plays the role of the cosmological constant because
$\dot{G}=0$; thus $w_{tot}\left(  Q_{1}\right)  =w_{tot}\left(  Q_{2}\right)
=-1.$ In order to study the stability of the point we apply the central
manifold theorem where we find that the family of solutions are stable for
values of $\lambda$ as they are given by the stability of point $P_{1}$.
\qquad

\item Point $Q_{2}~$has the same physical properties with $P_{2}$ and $\mu$ is
necessarily positive. However, the stability of the point is different; the
eigenvalues \ of the linearized system are calculated to $e_{1}\left(
Q_{2}\right)  =\sqrt{\frac{6}{\mu}},~e_{2}\left(  Q_{2}\right)  =6+\sqrt
{\frac{6}{\mu}}\left(  1-\lambda_{0}\right)  $; hence, $Q_{2}$ is a hyperbolic
(unstable) point, where the field $G\left(  t\right)  $ has a constant
equation of state parameter, i.e. $w_{G}\left(  Q_{2}\right)  =1-2\sqrt
{\frac{2}{3\mu}}$.

\item The eigenvalues for the linearized system close to the point $Q_{3}$ are
calculated to be $e_{1}\left(  Q_{3}\right)  =-\frac{1-\lambda_{0}}{\mu}~$and
$e_{2}\left(  Q_{3}\right)  =-3+\frac{\left(  1-\lambda_{0}\right)  ^{2}}%
{2\mu}$, which means that point $Q_{3}$ is stable when (a) $\lambda_{0}>1$ and
$\mu<0$ or (b) $\lambda_{0}<1$ and $\mu<\frac{\left(  1-\lambda_{0}\right)
^{2}}{6}$. As for the physical description of the solution at the point
$Q_{3}$, that is exactly the same as that of point $P_{3}$ for the power-law
potential. \ 

\item Point $Q_{4},~$describes a solution where potential $V\left(  G\right)
$ and the kinetic term of the field \thinspace$G$ contribute to the universe.
The equation of state parameter is calculated to be $w_{tot}\left(
Q_{4}\right)  =w_{G}\left(  Q_{4}\right)  =-1-\frac{1}{3\mu}$ which means that
it describes an accelerated universe for $\mu<-\frac{1}{2}$ or $\mu>0$. To
determine the stability of the point, we calculate eigenvalues which are
$e_{1}\left(  Q_{4}\right)  =-\frac{1}{\mu}$,~$e_{2}\left(  Q_{4}\right)
=-9+\frac{7}{2\mu}$. Hence, for $\mu>\frac{7}{18}$ both eigenvalues are
negative and the point is stable. Moreover, for $\mu>0~$someone can calculate
that $w_{tot}\left(  Q_{4}\right)  <-1$, which means that the parameter for
the total equation of state crosses the phantom divided line.

\item Point $Q_{5},~$can be seen as a special case of point $Q_{2}$~where
$\lambda_{0}=0$. The physical properties are the same as point $P_{1}$, that
is, $w_{tot}\left(  Q_{5}\right)  =w_{G}\left(  Q_{5}\right)  =1-2\sqrt
{\frac{2}{3\mu}}$. We calculate the eigenvalues of the linearized system, that
is, $e_{1}\left(  Q_{5}\right)  =-\sqrt{\frac{6}{\mu}}$,~$e_{2}\left(
Q_{5}\right)  =6+\sqrt{\frac{6}{\mu}}$, where we conclude that because
eigenvalue $e_{2}\left(  Q_{5}\right)  ~$has always a real positive value, the
solution which is described by point $Q_{5}$ is unstable.
\end{itemize}

\begin{figure}[ptb]
\includegraphics[height=6.5cm]{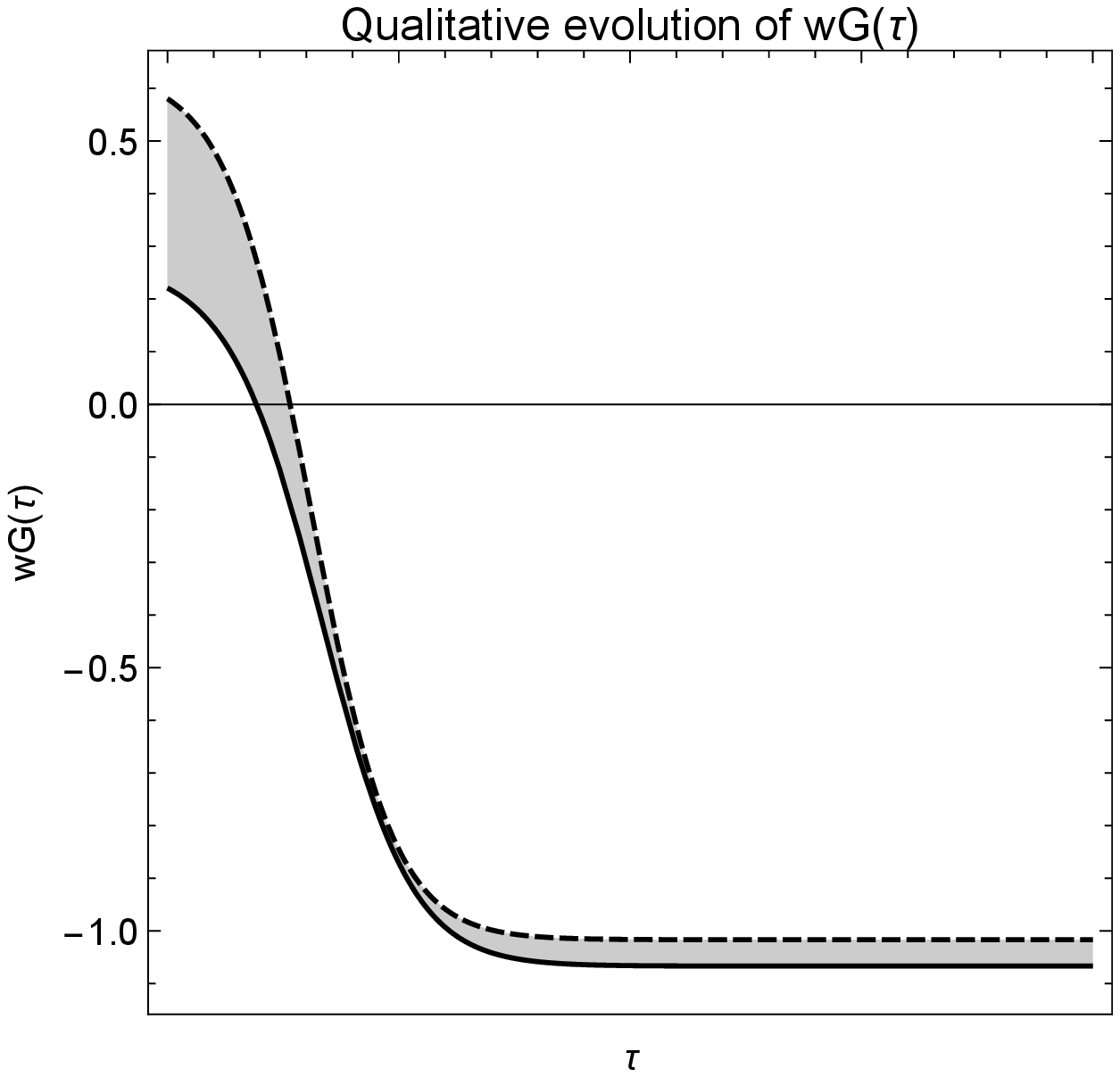}
\includegraphics[height=6.5cm]{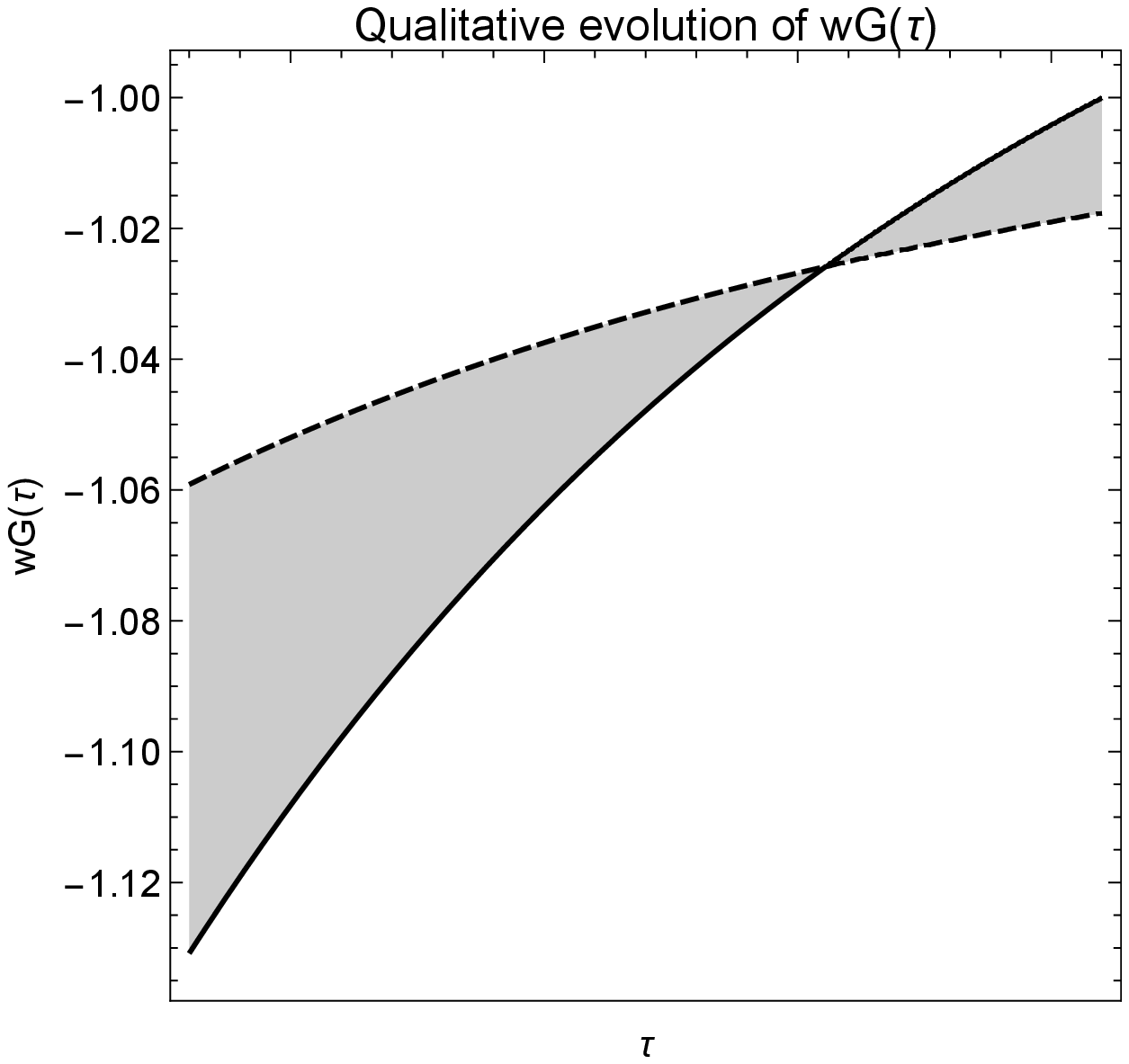}
\caption{Qualitative evolution for the equation of state parameter for various
values of the free parameter $\mu$, and for $\lambda_{0}=2.$ Left Fig. is for
$\mu>0,$ and specifically solid line is for $\mu=5,$ dashed line for $\mu=20.$
Right Fig. is for negative values of $\mu$ and in particular, solid line is
for $\mu=-5,$ dashed line for $\mu=-100.~$From the figure, it is clear that
positive values of $\mu$ are prefered in order for the $w_{G}$ parameter to be
free of singularities in the past.}%
\label{plott2}%
\end{figure}

Before we proceed to our analysis with the case in which we include matter
source, in Fig. \ref{plott2} we present the qualitative evolution of the
parameter for the equation of state $w_{G}\left(  \tau\right)  $, for positive
and negative values of the parameter $\mu$ and for $\lambda_{0}=2.$ For
positive values of $\mu$, the initial condition is for $x\left(  0\right)
\simeq1$, while we observe that the final attractor describes an accelerated
universe close to the de Sitter point. On the other hand, for negative values
of $\mu,$ i.e. $x<0$ and initial condition $x\left(  0\right)  \simeq-0.01$,
the final attractor is again close to the de Sitter universe. The value of the
parameter $\mu$ is unknown, and Fig. \ref{plott2} provides a qualitative
evolution of the equation of state parameter. \ From the numerical simulation,
we observe that the equation of state parameter can cross the phantom divine
line which does not contradict the observations \cite{planck2015}.

\subsection{Critical points with matter source}

As in the case of vacuum, we perform the same analysis for power-law and
exponential potential.

\subsubsection{Power-law potential}

For the power-law potential $V_{a}\left(  G\right)  $, where $\lambda
=const$.,~i.e. $\lambda=\lambda_{0}$, the dynamical (\ref{gl14}), (\ref{gl15})
admits the critical points~of the form $A=\left(  x\left(  A\right)  ,y\left(
A\right)  \right)  ,$
\[
A_{0}=\left(  0,0\right)  ~~,~~A_{1}=\left(  0,1\right)  ~~,~~A_{2}=\left(
1,0\right)  ~,
\]%
\[
~A_{3}=\left(  \frac{\left(  1-\lambda_{0}\right)  ^{2}}{6\mu},1-\frac{\left(
1-\lambda_{0}\right)  ^{2}}{6\mu}\right)  ~~,~~A_{4}=\left(  \frac{2}{3\mu
},0\right)  ~~,~~A_{5}=\left(  \frac{3\mu}{2\lambda_{0}^{2}},\frac
{2\lambda_{0}+3\mu}{2\lambda_{0}^{2}}\right)  \text{. }~
\]

\begin{figure}[ptb]
\includegraphics[height=6.5cm]{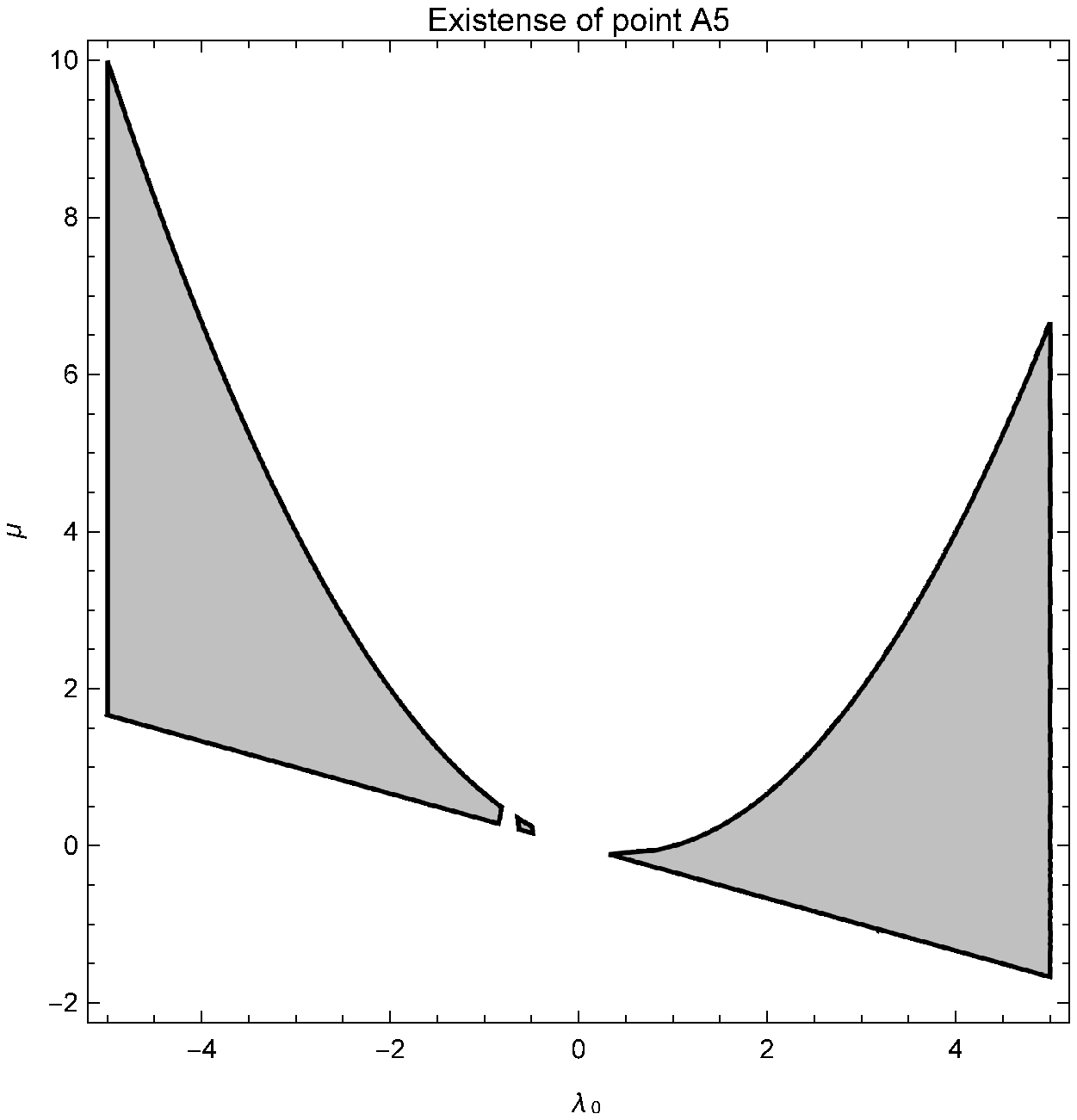}
\includegraphics[height=6.5cm]{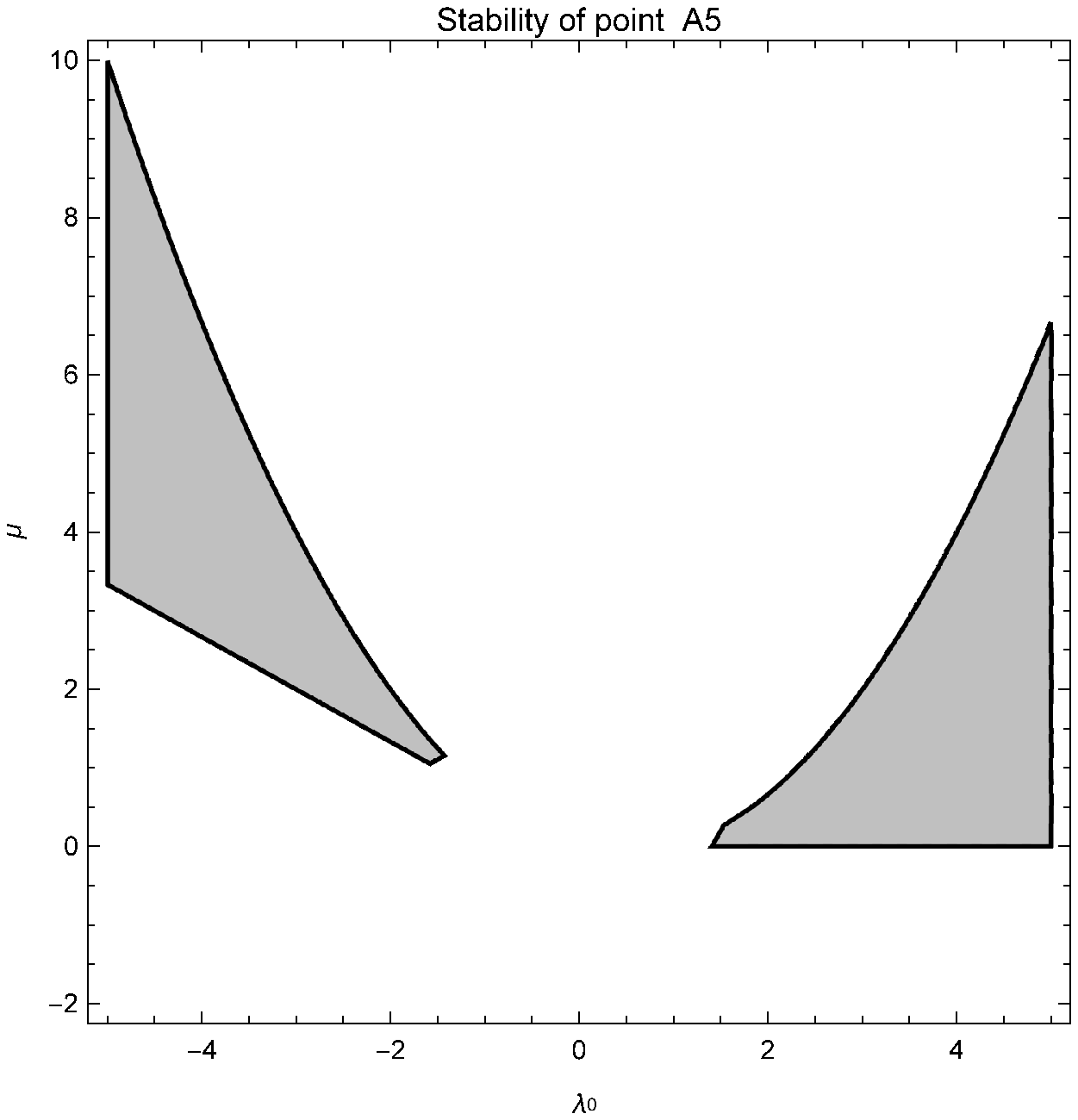}
\caption{Region plot on the space of the parameters $\lambda_{0}$ and $\mu$
where point $A_{5}$ exists (Left Fig.) and the point is stable\ (Right Fig.) }%
\label{plot33}%
\end{figure}

We observe that points $A_{1},~A_{2}$ and $A_{3}$ have the coordinates of
$P_{1},~P_{2}$ and $P_{3}$ respectively, while the new critical points are the
$A_{0},~A_{4}$ and $A_{5}$. \ More specifically for each critical point we have:

\begin{itemize}
\item Point $A_{0}$ corresponds to the matter dominated era where $\Omega
_{m}\left(  A_{0}\right)  =1$,$~\Omega_{G}\left(  A_{0}\right)  =0$ and
$w_{tot}=0.$ One of the eigenvalues of the linearized system close to the
critical point is positive which means that the point is unstable.

\item At the point $A_{1}$ the potential $V\left(  G\right)  $ dominates the
universe while $\dot{G}=0$, that is $\Omega_{m}\left(  A_{1}\right)  =0$ and
$\Omega_{G}\left(  A_{1}\right)  =1$, while $w_{tot}\left(  A_{1}\right)
=w_{G}\left(  A_{1}\right)  =-1$. The stability of the point is explicitly
that which is described for the point $P_{1}$.

\item The discussion of the physical properties for point $A_{2}$ is exactly
that for point $P_{2}$, because $\Omega_{m}\left(  A_{2}\right)  =0$ and
$\Omega_{G}\left(  A_{2}\right)  =1.$ However, the eigenvalues are calculated
to be $e_{1}\left(  A_{2}\right)  =3+\sqrt{\frac{6}{\mu}},$ and $e_{2}\left(
A_{2}\right)  =6+\left(  1-\lambda_{0}\right)  \sqrt{\frac{6}{\mu}}$, that is,
the point is always unstable because $\operatorname{Re}\left(  e_{1}\left(
A_{2}\right)  \right)  >0$.

\item Point $A_{3}$ exists for all the values $\lambda$ and $\mu,$ while the
physical solution is that described by point $P_{3}$. The eigenvalues of the
point are calculated to be $e_{1}\left(  A_{3}\right)  =-3+\frac{\left(
1-\lambda_{0}\right)  ^{2}}{2\mu}$,~and $e_{2}\left(  A_{3}\right)
=-3-\frac{\left(  1-\lambda_{0}\right)  \lambda_{0}}{\mu}$. The stability of
the point depends on the values of the parameters $\lambda_{0},~\mu~$and
specifically the eigenvalues are positive, that is,~$A_{3}$ is unstable when
(a) $\lambda_{0}\leq-1~$or $\lambda_{0}>1~$and $0<\mu<\frac{\left(
1-\lambda_{0}\right)  ^{2}}{6}$; (b) $-1<\lambda_{0}<0$ and $0<\mu
<\frac{\lambda_{0}\left(  \lambda_{0}-1\right)  }{3}$.

\item Point $A_{4}$ describes a universe where $\Omega_{m}\left(
A_{4}\right)  =1-\frac{2}{3\mu}$ and $\Omega_{G}\left(  A_{4}\right)
=\frac{2}{3\mu}$; while for the parameters of the equation of state
$w_{tot}\left(  A_{4}\right)  =-\frac{2}{3\mu}$ and $w_{G}\left(
A_{4}\right)  =-1~$follows. Hence, field $G$ behaves like a cosmological
constant and, specifically, that point corresponds to the $\Lambda$CDM
cosmology, where parameter $\mu$ is related to the energy density of the dark
energy. \ It is important to mention that the point exists only for values of
$\mu$ where $0\leq\Omega_{m}\left(  A_{4}\right)  \leq1$, that is, $\mu
\geq\frac{2}{3}$. The corresponding eigenvalues of the linearized system are
$e_{1}\left(  A_{4}\right)  =-\frac{9}{2}+\frac{7}{\mu}$,~$e_{2}\left(
A_{4}\right)  =-3+\frac{2\left(  2-\lambda_{0}\right)  }{\mu};$ thus, we
conclude that point $A_{4}$ is stable for $\lambda>\frac{13}{3}$ and
$\frac{14}{9}<\mu<\frac{2\left(  \lambda_{0}-2\right)  }{3}$.~However, when
point $A_{4}$ is stable, we calculate $\Omega_{m}\left(  A_{4}\right)
>\frac{4}{7}$, which is bigger value from the observable one, i.e.
$\Omega_{m0}\simeq0.28$. Therefore, in order for the model to be
cosmologically viable, point $A_{4}$ has to be unstable.

\item Point $A_{5}$ exists when (a) $\lambda_{0}<0$ with $-\frac{\lambda_{0}%
}{3}<\mu<\frac{\lambda_{0}\left(  \lambda_{0}-1\right)  }{3}$; (b)$~\lambda
_{0}>0$ with~$-\frac{\lambda_{0}}{3}<\mu<\frac{\lambda_{0}\left(  \lambda
_{0}-1\right)  }{3},$these ranges are given in Fig. \ref{plot33}. Point
$A_{5}$ describes a universe where $\Omega_{m}\left(  A_{5}\right)
=1+\frac{1}{\lambda_{0}}\left(  \frac{3\mu}{\lambda_{0}}-1\right)  $; and
$\Omega_{G}\left(  A_{5}\right)  =\frac{1}{\lambda_{0}}\left(  \frac{3\mu
}{\lambda_{0}}-1\right)  $ with equation of state parameters $w_{G}\left(
A_{5}\right)  =-\frac{3\lambda_{0}}{3\mu+\lambda_{0}}$ and $w_{tot}\left(
A_{5}\right)  =-\frac{3}{\lambda_{0}}$, where $w_{tot}\left(  A_{5}\right)
<-\frac{1}{3}$ for $0<\lambda_{0}<9$. \ The point describes a universe with
radiation and dark matter when $\mu=-\frac{10\lambda_{0}}{3}$ and $\lambda
_{0}\leq9$. It is an interesting point because it can describe a phase where
radiation dominates the universe for $\lambda\rightarrow9^{+}$. \ Point
$A_{5}$ is stable when (i)~$\lambda_{0}<-1$ with $-\frac{2\lambda_{0}}{3}%
<\mu<\frac{\lambda_{0}\left(  \lambda_{0}-1\right)  }{3}$ \ and (ii)
$\lambda_{0}>1$, $0<\mu<\frac{\lambda_{0}\left(  \lambda_{0}-1\right)  }{3}$
as they are given in Fig. \ref{plot33}.
\end{itemize}

We continue our analysis with the scenario of the exponential potential
$V_{\left(  b\right)  }\left(  G\right)  $.

\subsubsection{Exponential Potential}

Consider now the exponential potential $V_{\left(  b\right)  }\left(
G\right)  $. The dynamical system (\ref{gl14})-(\ref{gl16}) has dimension
three and the critical points are of the form$~B=B\left(  x\left(  B\right)
,y\left(  B\right)  ,\lambda\left(  B\right)  \right)  $, in particular%
\[
B_{0}=\left(  0,0,\lambda\right)  ~~,~~B_{1}=\left(  0,1,\lambda\right)
~~,~~B_{2}=\left(  1,0,\lambda_{0}\right)  ~,
\]%
\[
B_{3}=\left(  \frac{\left(  1-\lambda_{0}\right)  ^{2}}{6\mu},1-\frac{\left(
1-\lambda_{0}\right)  ^{2}}{6\mu},\lambda_{0}\right)  ~~,~~B_{4}=\left(
\frac{2}{3\mu},0,\lambda_{0}\right)  ~~,~~B_{5}=\left(  \frac{3\mu}%
{2\lambda^{2}},\frac{2\lambda+3\mu}{2\lambda^{2}},\lambda_{0}\right)  ,
\]%
\[
B_{6}=\left(  \frac{1}{6\mu},1-\frac{1}{6\mu},0\right)  ~~,~~B_{7}=\left(
1,0,0\right)  ~~,~~B_{8}=\left(  \frac{2}{3\mu},0,0\right)  \text{.}%
\]

Points $B_{0}-B_{5}$ are specifically points $A_{0}-A_{5}$ respectively, while
$B_{6}$ and $B_{7}$ are related to $Q_{4}$ and $Q_{5}$ in the vacuum scenario,
and $B_{8}$ is the only new point which is a special of point $A_{4}$ with
$\lambda$ zero. Because of that correspondence, it is not necessary to discuss
the physical properties of the points; therefore, we continue with the
discussion of the stability conditions.

\begin{itemize}
\item Point $B_{0}$ is always unstable because one of the eigenvalues is
always positive.

\item Point $B_{1}$ has two zero eigenvalues, hence central manifold theorem
has to be applied. In particular, the coordinates of $B_{1}$ describe a line
in the space $\left\{  x,y,\lambda\right\}  $. We find that the stability and
instability of the solution corresponds explicitly to the conditions given by
point $P_{1}.$

\item Point $B_{2}$ has the eigenvalues $e_{1}\left(  B_{2}\right)
=3+\sqrt{\frac{6}{\mu}},~e_{2}\left(  B_{2}\right)  =3+\sqrt{\frac{6}{\mu}%
}\left(  1-\lambda\right)  $ and $e_{3}\left(  B_{3}\right)  =\sqrt{\frac
{6}{\mu}}$; hence the point is always unstable.

\item Point $B_{3}$ is found to be stable when parameters $\lambda_{0}$ and
$\mu$ are given by the following set of ranges: (a) For $\lambda\leq-1$%
,~$\mu>\frac{\lambda_{0}\left(  \lambda_{0}-1\right)  }{3}$; (b) for
$-1<\lambda_{0}<1,~\mu>\frac{\left(  1-\lambda_{0}\right)  ^{2}}{6},$ and (c)
for $\lambda>1,$ $\mu<0.$ The surface in the space of variables $\left\{
\lambda_{0},\mu\right\}  $ in which point $B_{3}$ is stable is presented in
Fig. \ref{plotb3}. \begin{figure}[ptb]
\includegraphics[height=6.5cm]{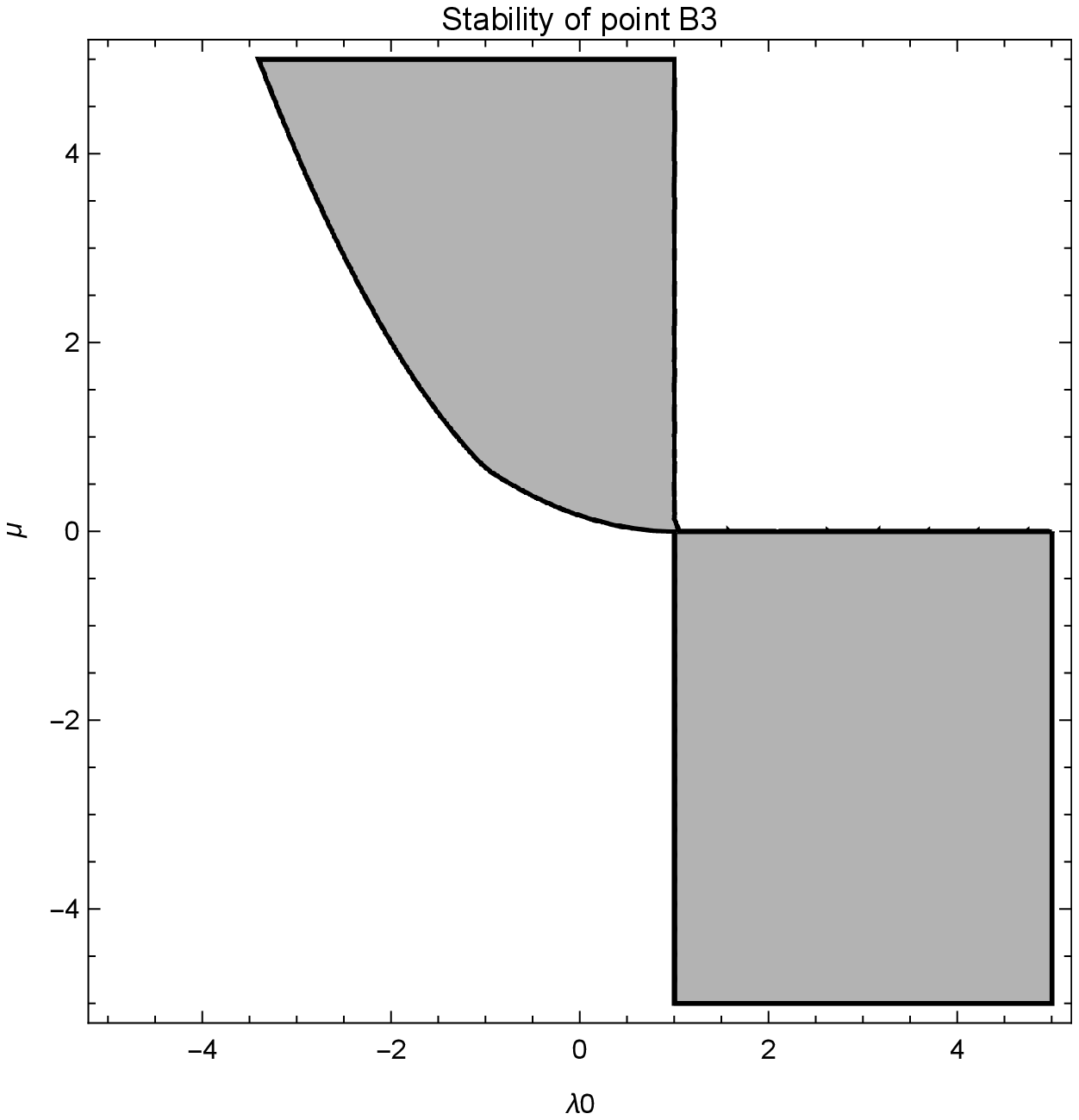}
\includegraphics[height=6.5cm]{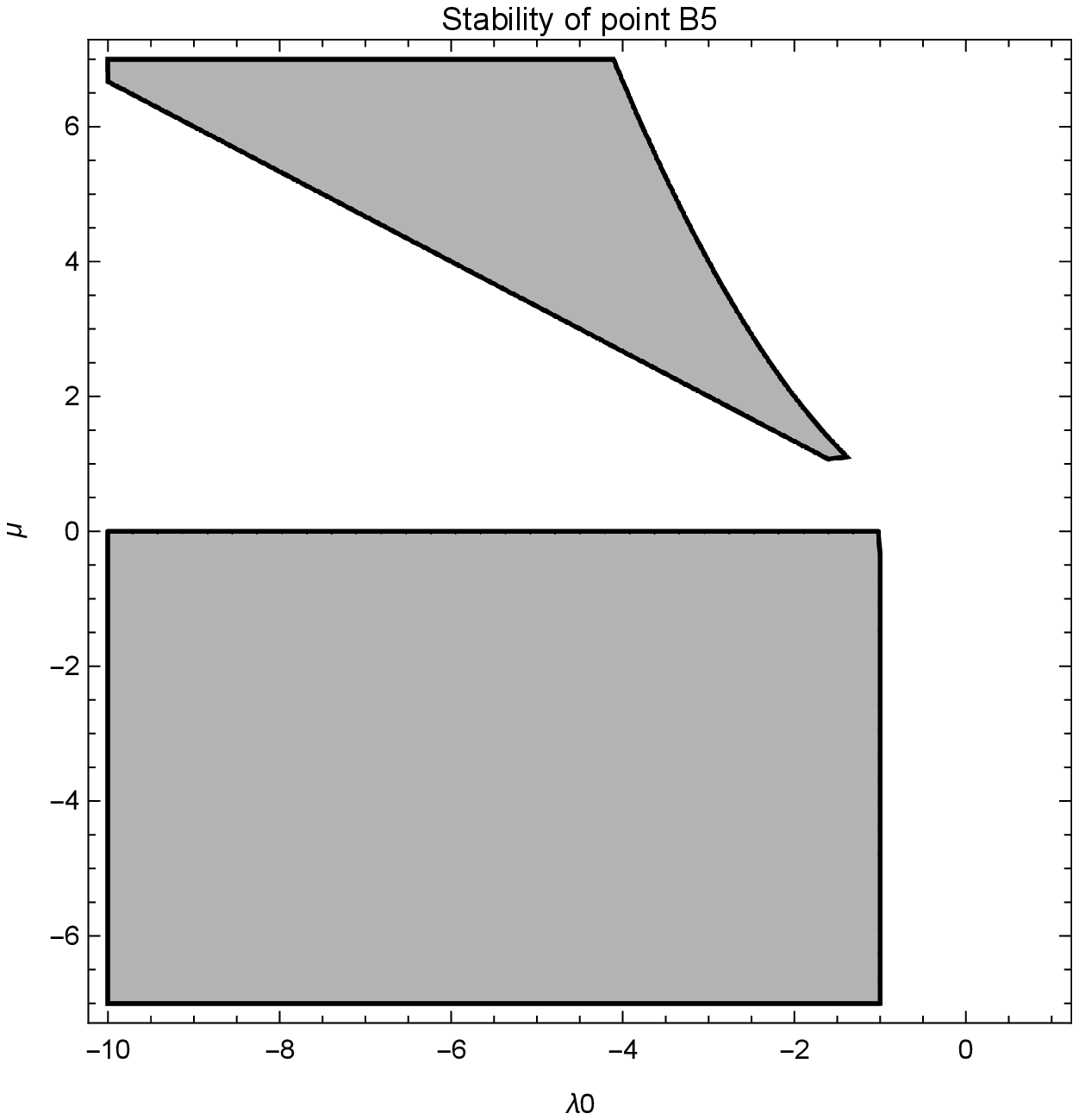}
\caption{Region plot on the space of the parameters $\lambda_{0}$ and $\mu$
where points $B_{3}$ (Left. Fig) and $B_{5}$ (Right Fig.) are stable.}%
\label{plotb3}%
\end{figure}

\item At the point $B_{4},$ the eigenvalues of the linearized system have the
simple expressions, $e_{1}\left(  B_{4}\right)  =\frac{2}{\mu},~e_{2}\left(
B_{4}\right)  =-\frac{9}{2}+\frac{7}{\mu}$ and $e_{3}\left(  B_{4}\right)
=3+\frac{2\left(  2-\lambda_{0}\right)  }{\mu}$, which means that the point is
stable when $\lambda_{0}<0$ and $2\lambda_{0}-4<3\mu<0$.

\item Point $B_{5}$ is stable when (a) $\lambda_{0}<-1$ and $\mu<0$ or
$-\frac{2\lambda_{0}}{3}\mu<\frac{\lambda_{0}\left(  \lambda_{0}-1\right)
}{3}$ as it is presented in Fig. \ref{plotb3}.

\item Point $B_{6}$ provides the eigenvalues $e_{1}\left(  B_{6}\right)
=-\sqrt{\frac{6}{\mu}},~$\ $e_{2}\left(  B_{6}\right)  =3+\sqrt{\frac{6}{\mu}%
}$ and $e_{3}=6+\sqrt{\frac{6}{\mu}}$ which means that the solution at the
point is always unstable.

\item Close to point $B_{7}$ the eigenvalues of the linearized system are
$e_{1}\left(  B_{7}\right)  =-\frac{1}{\mu},~e_{2}\left(  B_{7}\right)
=-6+\frac{7}{2\mu}$ and $e_{3}\left(  B_{7}\right)  =-3+\frac{2}{3\mu}$, from
where it follows that the point is always stable for every value of $\mu
>\frac{2}{3}$.

\item The eigenvalues at point $B_{8}$ are derived to be $e_{1}\left(
B_{8}\right)  =-\frac{2}{\mu},~e_{2}\left(  B_{8}\right)  =3+\frac{4}{\mu}$
and $e_{3}\left(  B_{8}\right)  =-\frac{9}{2}+\frac{7}{\mu},$ which means that
the point is always unstable.
\end{itemize}

It is important to mention that our study for the power-law and the
exponential potentials coverS all the possible physical states which can be
determined by the dynamical system\ (\ref{gl14})-(\ref{gl16}). The only
differences will be on the stability of the points. Therefore, it is not
necessary to extend the present analysis for other kind of potentials.

In order to explain the latter statement, we not that any stationary point
corresponds to a value $\lambda_{0}$ such that $\Gamma\left(  \lambda
_{0}\right)  =const.$ Now we can always rescale a new variable $\bar{\lambda
}_{0}$, such that these points to be described by the exponential potential.
For instance, consider the hyperbolic potential for the minimally coupled
scalar field studied in \cite{bhyper,chyper,dhyper}. The admitted critical
points \cite{ahyper} correspond to eras where the hyperbolic potential mimics
the exponential potential or the power-law potential \cite{amen}.

\section{Conclusions}

\label{con}

In this work, we applied the method of fixed point analysis in order to study
the cosmological viability of a gravitational theory with varying $G$ and
$\Lambda$, which was proposed in \cite{alfio1}. In the renormalization group
approach, there is not a unique way to perform the modification of the
fundamental \textquotedblleft constants\textquotedblright. In \cite{alfio1}
the authors proposed the modification to be done in the ADM Lagrangian, which
leads to the introduction of a field $G$ different from that of the
scalar-tensor theories. On the other hand, as it has been found in
\cite{ref1}, the modification of $G$ and $\Lambda$ in Einstein-Hilbert Action
can lead to Brans-Dicke like gravitational theory. Another equivalent way to
reproduce the field equations of \cite{alfio1} is the renormalization group to
be applied in field equation's of Einstein's General Relativity.

In the cosmological scenario of a spatially flat FLRW universe, the resulting
field equations are of second-order with free variables the scale factor
$a\left(  t\right)  $ and the field $G\left(  t\right)  $, where the
cosmological constant plays the role of the potential for the field~$G$, that
is, we considered $\Lambda\left(  t\right)  =\Lambda\left(  G\left(  t\right)
\right)  $. Furthermore, in our cosmological scenario, minimally coupled
pressureless matter source has been introduced.

In order to perform the dynamical analysis, we define new dimensionless
variables while the field equations were rewritten as an
algebraic-differential system consisted by three first-order differential
equations. For two exact forms of the \textquotedblleft
potential\textquotedblright\ term $\Lambda\left(  G\left(  t\right)  \right)
$ the critical/fixed points for the reduced system of algebraic-differential
equations ARE determined. The exact forms of the potentials that we selected
cover all the possible different families of points with the same physical
properties, which can be provided by the theory for any other form of the
\textquotedblleft potential\textquotedblright\ $\Lambda\left(  G\left(
t\right)  \right)  $.

For the vacuum scenario and for power-law potential, we determined three
critical points. Two of the points, namely $P_{2}$ and $P_{3}$, provide (in
general) power-law scale factors corresponding to ideal gas solutions while
the physical solution for the third point, $P_{1}$, describes a de Sitter
universe. For the exponential potential, in addition to the above, two new
critical points are determined, $Q_{4}$ and $Q_{5},$ which describe singular
solutions of the form $a\left(  t\right)  =a_{0}t^{\kappa}$, with
$\kappa=\kappa\left(  \mu,\lambda\right)  $.

In the presence of matter, new critical points are determined, where the
matter source contributes to the final state of the universe. For the power
law potential, the points with the new physical solutions are the $A_{0}%
$,$~A_{4}~$and $A_{5}$. Point $A_{0}$ describes the matter dominated era where
$\Omega_{m}=1$ while the solution for the scale factor $a\left(  t\right)
=a_{0}t^{\frac{2}{3}}$ is always unstable. On the other hand, at the points
$A_{4},~A_{5}$, the field $G$ and the pressureless matter contribute to the
evolution of the universe, that is $\Omega_{m}\left(  \mu,\lambda\right)
\neq0$, and $\Omega_{G}\left(  \mu,\lambda\right)  \neq0$. At point $A_{4}$
the parameter for the equation of state has value $-1$, which means that IT
mimics the cosmological constant and the point describes the limit of the
$\Lambda$CDM universe. However, in order for the point to be physically
accepted and to be in comparison with the observations, it has to be unstable.
Moreover, at point $A_{5}$, field $G$ acts as an ideal gas and it is possible
to describe en epoch with radiation and matter sources.

\begin{figure}[ptb]
\includegraphics[height=6.5cm]{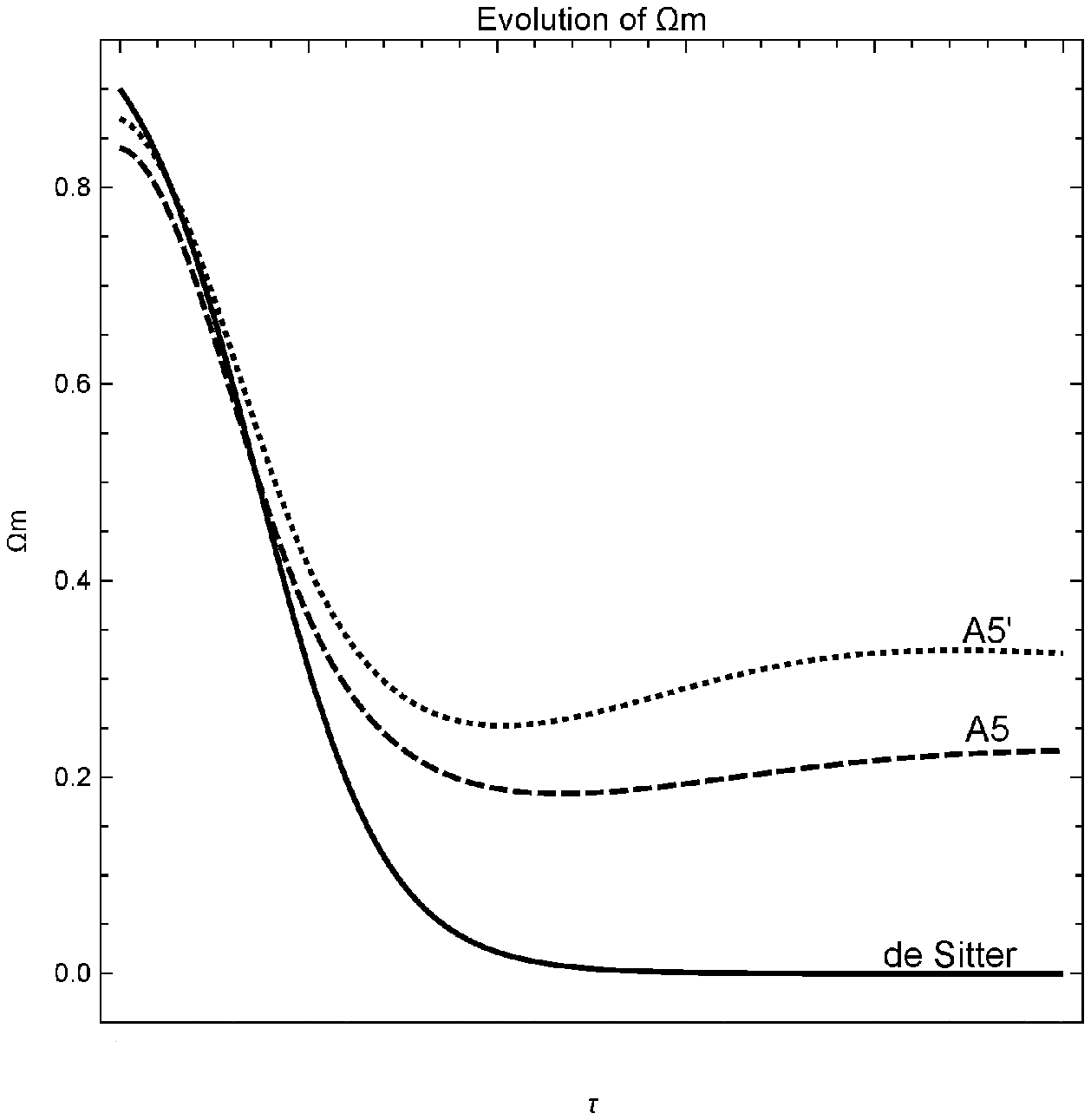}
\includegraphics[height=6.5cm]{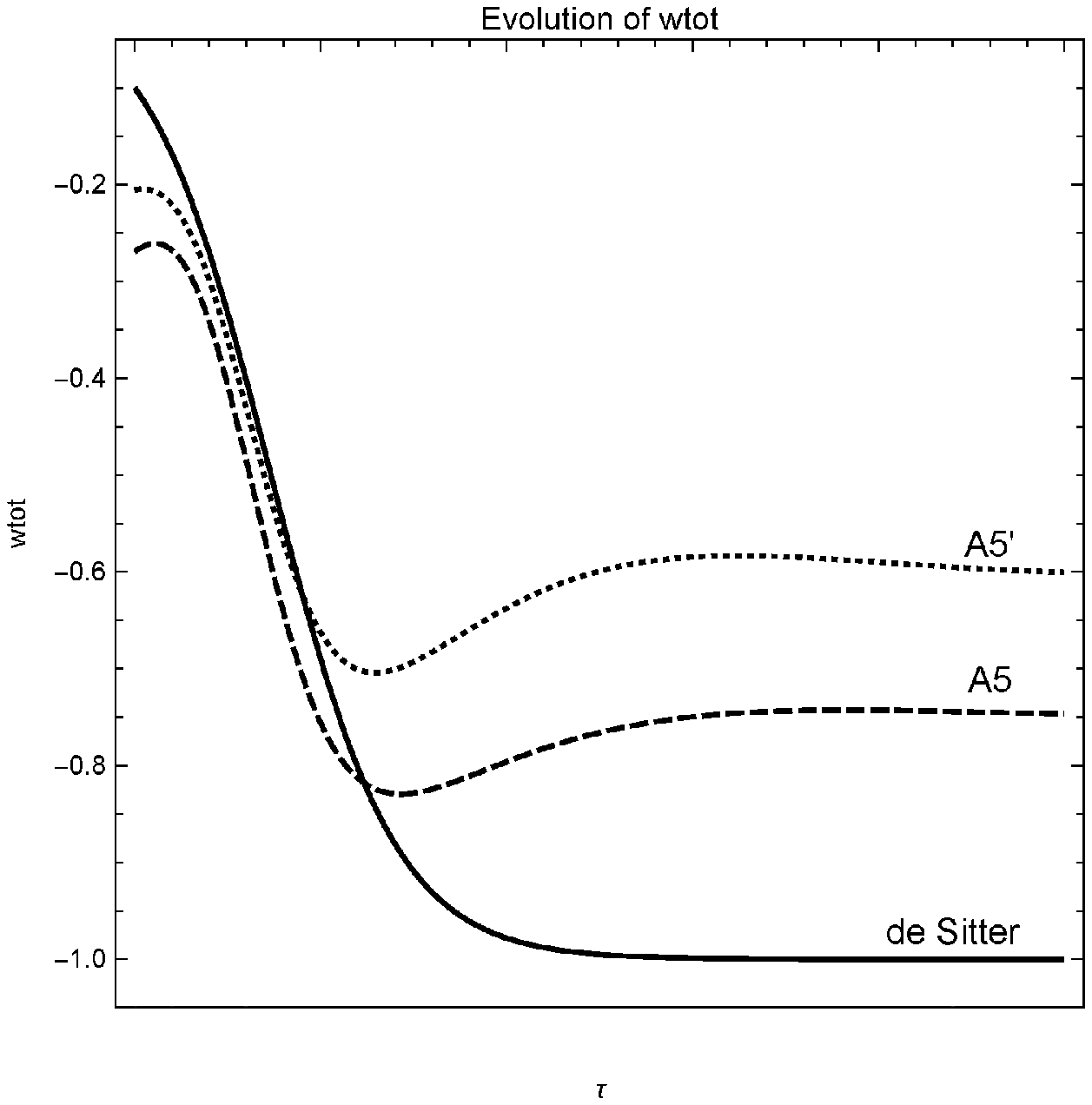}
\caption{Qualitative evolution of the energy density $\Omega_{m}$ and the
equation of state parameter $w_{tot}$ for initial conditions close to the
point $A_{0}$, for the power-law potential $V_{\left(  a\right)  }\left(
G\right)  $. \ Solid lines are for initial condition with $x\left(  0\right)
=0$, and for values$~\left(  \lambda_{0},\mu\right)  ~=\left(  3,1\right)  ,$
where it is clear that the final state of the universe is the de Sitter
solution. The dashed and the dotted lines are for initial conditions with
$x\left(  0\right)  \neq0$, and for the set of variables $\left(  \lambda
_{0},\mu\right)  =\left(  4,2.8\right)  $ (dashed), $\left(  \lambda_{0}%
,\mu\right)  =\left(  5,4\right)  $ dotted. The free parameters have been
chosen such point $A_{5}$ to be the attractor in the evolution of the
universe.}%
\label{figg1}%
\end{figure}

Numerical simulations for the evolution of the energy density parameter
$\Omega_{m}$ and the equation of state parameter $w_{tot}$ are presented in
Fig. \ref{figg1} for initial conditions close to the point $A_{0}$, for
different values of the parameters $\lambda$ and $\mu$ such that point $A_{5}$
is an attractor.

Finally, in the case of the exponential potential, only one extra point was
found, namely $B_{7}$, (including those listed above) which has the same
physical properties with point $A_{4}\,.$ However, stability analysis provides
that the solution at point $B_{7}$ is always unstable. 

From our analysis, it is clear that the theory provides the basic cosmological
eras in the evolution of the universe. However, there are differences with
other second-order theories, such as the scalar-tensor theories. In
particular, the role of the interaction parameter $\mu$ is actually unknown
but we can see that it can be related to the energy density $\Omega_{G}$ as
some of the critical points, while from our results, it is clear that its
possible values can be demanding the the existence and stability of some
specific critical points.

There are various similarities of the critical points with that of Brans-Dicke
theory \cite{dn8,papag}. For instance, in the case of vacuum and for a
power-law potential, both theories admit three critical points \cite{papag}
while the physical properties of the critical points/solutions depend on the
Brans-Dicke parameter or parameter $\mu$ respectively. However, while the
theory of our consideration always admits the de Sitter universe (point
$P_{1}$) as a critical point for arbitrary power-law potential, for the
Brans-Dicke theory that is true, if and only if, the power-law potential is
the quadratic. Other differences between the two theories appear when we
include matter source, or generalize the form of the potential.

Consider the coordinate transformation%
\begin{equation}
a\rightarrow A\phi~,~G\rightarrow\phi^{2}\label{pp.01}%
\end{equation}
Hence, Lagrangian (\ref{gl4}) becomes%
\begin{equation}
\mathcal{L}\left(  a,\dot{a},\phi,\dot{\phi}\right)  =\frac{1}{N}\left(
-3A\phi\dot{A}^{2}-3A^{2}\dot{A}\dot{\phi}+\frac{1}{2}\frac{2\mu-6}{\phi}%
A^{3}\dot{\phi}^{2}\right)  -Na^{3}\phi^{3}V\left(  \phi\right)  \label{pp.02}%
\end{equation}
which is the Lagrangian describes the field equations for the Brans-Dicke
theory%
\begin{equation}
S=\int dx^{4}\sqrt{-g}\left[  \frac{1}{2}\phi R-\frac{1}{2}\frac{\omega_{BD}%
}{\phi}g^{\mu\nu}\phi_{;\mu}\phi_{;\nu}-V_{BD}\left(  \phi\right)  \right]
,\label{pp.03}%
\end{equation}
for the line element%
\begin{equation}
ds^{2}=-N^{2}dt^{2}+A^{2}\left(  t\right)  \left(  dx^{2}+dy^{2}%
+dz^{2}\right)  \label{pp.04}%
\end{equation}
where $\omega_{BD}=2\mu-6$ and $V_{BD}=\phi^{3}V\left(  \phi\right)  $.
\ Recall that transformation (\ref{pp.01}) is not a conformal transformation,
consequently the two Lagrangians (\ref{gl4}), (\ref{pp.02}) are not conformal
equivalents, it is just the same Lagrangian in different coordinates. However,
these two point-like Lagrangians describe the field equations for two
different gravitational theories for the line elements (\ref{gl3})
and(\ref{pp.04}). Transformation (\ref{pp.01}) is important because we can
transform solutions of one theory into solutions of the other theory. Another
important observation is that when $\mu=3$, Lagrangian (\ref{pp.02}) describes
the gravitational field equations of $f\left(  R\right)  $-gravity, for
details see \cite{paliafr1} and references therein. 

Without loss of generality we select $N\left(  t\right)  =1$; then for the
power law potential $V\left(  G\right)  =V_{0}G^{Q}$ in (\ref{gl4}) and in the
case of vacuum, i.e. $\rho_{m0}=0$, we determine the exact solution for the
varying $G$ and $\Lambda$ theory%
\begin{equation}
a\left(  t\right)  =a_{0}t^{\frac{2\mu}{Q-1}}~,~~G\left(  t\right)
=G_{0}t^{-\frac{2}{1+Q}}%
\end{equation}
with $V_{0}=\left(  1+Q\right)  ^{-2}G_{0}^{-1-Q}\left(  \left(  1-Q\right)
^{2}-6\mu\right)  \mu.~$The latter solution describes a perfect fluid solution
with equation of state parameter $w_{G\left(  t\right)  }=-1+\frac{Q-3\mu
}{3\mu}.$

Moreover, under the coordinate transformation (\ref{pp.01}) we find the
Brans-Dicke equivalent potential to be $V_{BD}\left(  \phi\right)  =V_{0}%
\phi^{3+2Q}$, while the exact solution becomes%
\begin{equation}
A\left(  t\right)  =a_{0}t^{\frac{2\mu\left(  1+Q\right)  +\left(  Q-1\right)
}{Q^{2}-1}}~~,~~\phi\left(  t\right)  =G_{0}t^{-\frac{1}{1+Q}}%
\end{equation}
which corresponds to a perfect fluid solution with equation of state parameter
$w_{BD}=-\frac{6\mu\left(  1+Q\right)  +3Q-\left(  1+2Q^{2}\right)  }{3\left(
2\mu\left(  1+Q\right)  +\left(  Q-1\right)  \right)  }.$

Hence, in order to see the differences between the two solutions we set
$\mu=1$, where we find that $\left\vert w_{G\left(  t\right)  }\right\vert <1$
for $1<Q<7$, while for the Brans-Dicke solution we determine that $\left\vert
w_{BD}\right\vert <1$ when $-1<Q<\frac{9}{2}-\frac{\sqrt{97}}{2}$ and
$1<Q<\frac{9}{2}+\frac{\sqrt{97}}{2}$.

A more detailed analysis and comparison with cosmological data are necessary
in order for the role of parameter $\mu$ to be determined. Such an analysis
extends the scope of this work and will be published elsewhere.

\begin{acknowledgments}
AP acknowledges the financial support of FONDECYT grant no. 3160121 and thanks
the University of Athens and the $KE\Pi B/\Sigma\Lambda$ for the hospitality
provided while part of this work was performed.
\end{acknowledgments}

\end{document}